quant-ph/0407180

\documentclass[pra,12pt,eqsecnum]{revtex4}
\usepackage{graphicx}

\begin{document}

\def\BE{\begin{equation}}
\def\EE{\end{equation}}
\def\BY{\begin{eqnarray}}
\def\EY{\end{eqnarray}}
\def\BA{\begin{array}}
\def\EA{\end{array}}

\def\L{\label}

\def\({\left(}
\def\){\right)}
\def\[{\left[}
\def\]{\right]}

\def\o{\overline}
\def\nn{\nonumber}

\title{Controlling a group velocity of light by magnetic field}
\author{ Yu. M. Golubev and T. Yu. Golubeva }
\affiliation{V. A. Fock Physics Institute, St. Petersburg State University, ul.
Ul'anovskaya 1, 198504 St. Petersburg, Stary Petershof, Russia \\}
\author{ Yu.~V.~Rostovtsev
and M. O. Scully }
 \affiliation{Department of Physics and Institute for
Quantum Studies, Texas A \& M University, College Station, 77843, USA}
\date{\today}

\begin{abstract}
\vskip12pt

We have shown that quantum interference in a driven quasi-degenerate two-level
atomic system can be controlled by an externally applied magnetic field. We
demonstrate that the mechanism of optical control is based on quantum
interference, which allows one to implement both electromagnetically induced
transparency and electromagnetically induced absorption in one atomic system.
Dispersion of such the medium allows one to control group velocity of
propagation of light pulses be ultra-slow or superluminal via applied magnetic
field.

\end{abstract}
\maketitle

\section{Introduction}
\L{I} Quantum coherence and interference play an important role in the
interaction of coherent laser fields with atomic systems. As has been shown that
interplay between destructive and constructive interference of atomic
transitions \cite{1} leads to electromagnetically induced transparency (EIT) or
electromagnetically induced absorption (EIA)~\cite{2}. EIT has found a wide
variety of applications in quantum optics and nonlinear optical processes
\cite{3}. EIA could have potential applications to high-speed optical modulation
and quantum switching \cite{4,5}.

$N$-type scheme has recently attracted much attention and opened a new approach
to manipulate the nonlinear phenomena in optical process \cite{6}. It has been
shown that atomic coherence among Zeeman sublevels can be spontaneously
transferred from upper level to lower one, which gives rise not only to EIT, but
also to EIA \cite{2,7,8}. Doppler-free resonance absorption observed in the $N$
scheme displays another interesting features in the Doppler-broadened
medium~\cite{9}. Three-photon Doppler-free resonance  has been observed in hot
Rb vapor driven with one of two coherent fields far detuning from its resonance
in the $N$ scheme\cite{10}. The other type of schemes have been analysed in
Refs. \cite{11,12}.

Usually, the atomic system has one type of interference, either
constructive or destructive, that depends on the configuration of
atomic levels and laser fields.
Recently, we show a way to coherently control the type of interference in one
quasi-degenerate two-level atomic system by applying an external magnetic field
\cite{15}.

In this paper we show a possibility of control by means of magnetic field not
only of  the type of interference but of group velocity too. Application of
magnetic field can change velocity of propagation of light pulses from
ultra-slow to superluminal.

The paper is organized as follows. In Sec.~\ref{II} the physical model and  the
main equations are presented.  The equations are solved in the stationary
conditions.  In Sec.~\ref{III} together with App.~\ref{App1} the coherency  is
given for the thermal resonant medium. In Sec.~\ref{IV} the case of the
immovable atoms in the N-cofiguration is investigated.  At last, in Sec.~\ref{V}
together with App.~\ref{App2} the EIT, EIA and slow light effects are
investigated in the thermal vapor.

\section{The physical model and the basic equations}\L{II}

In Figure\ 1 the experimental setup discussed below is shown. It is assumed a
light beam, consisting of two monochromatic waves, which are propagated into the
same direction, with frequencies $ \omega $ (the drive wave) and $ \omega_p $
(the probe wave), passes through a cell of an atomic vapor at room temperature.
It is assumed that each atom, participating in the thermal motion, crosses the
light beam and after that looses  any information about past interaction with
light on the walls of cell. It means that any atom goes in into the light beam
volume in a ground state ever independently of for the first time or otherwise.

Let  the atom has effectively a two-level energetic structure with a twice
degeneration of both the levels (see fig.\ 2).  In an outside magnetic field the
Zeeman's effect takes a place, and generally speaking it leads to respective
splitting the levels. Let  some conditions (for example, the light
polarizations) ensure interactions of the drive field  with the atom on both the
$ (ac) $- and $ (bd) $- transitions and of the probe wave - on the  $ (ab)
$-transition.

We take into account incoherent processes in the atomic system.
First, there are
 spontaneous emissions on the  $(ab)$- and
$(ac)$-transitions with rates equaled  to $\gamma$ and on the $(dc)$- transition
- with $2\gamma$.  Second, we model processes of arrival of atoms into the light
beam volume  and departure  of them from there as incoherent processes too. As
is seen in Fig.\ 2 there are a "pump"  with mean rate $\mu$ to the lower
sub-levels and an additional "decay" of all the sub-levels with rate $\gamma_0$.

We start with an  equation for an atomic density matrix $\hat\sigma$ in the
interaction picture for the single four-level atom, interacting with two
coherent classical fields with complex amplitudes $\alpha$ (the drive field) and
$\alpha_p$ (the probe field):
\BY
&&\dot {\hat\sigma} =-i \[\hat V, \hat\sigma \]-\hat R\hat\sigma +\hat
\mu\L{2.1}.
\EY
The  $\hat R$ and $\hat\mu$ operators ensure the mentioned above incoherent
processes (see Fig.\ 2). The atom-field interaction Hamiltonian has a well known
form:
\BY
&&\hat V =-g\alpha\mid a\rangle\langle c\mid e^ {i\Delta_1 t} -g\alpha\mid
d\rangle\langle b\mid e^ {i\Delta_2 t}-g_p\alpha_p\mid a\rangle\langle b\mid e^
{i\Delta_p t} +h.c.\L{2.2}
 \EY
  Here are frequency detunings:  $ \Delta_1
=\omega_{ac}-\omega,\; \Delta_2 =\omega_{db}-\omega,\;\Delta_p
=\omega_{ab}-\omega_p $ ($\omega_{ac},\;\omega_{bd}$ and $\omega_{ab}$ - the
respective atomic frequencies).

Further we will use not the field amplitudes but  the so-called Rabi-frequencies
$ \Omega=g\alpha $ and $ \Omega_p=g_p\alpha_p $.

 Rewriting the equation (\ref{2.1}) in terms of matrix elements, we can obtain
  the following system of equations:
\BY &&\dot\sigma_{aa} =-\(2\gamma+\gamma_0\)\sigma_{aa} +
i\Omega_p\sigma_{ba}-i\Omega_p^*\sigma_{ab} +
i\Omega\sigma_{ca}-i\Omega^*\sigma_{ac} =0\L {2.3} \\ &&\dot\sigma_{bb} = \(\mu
+\gamma\sigma_{aa} \)-\gamma_0\sigma_{bb} +2\gamma\sigma_{dd}-
i\Omega_p\sigma_{ba} + i\Omega_p^*\sigma_{ab} +
i\Omega^*\sigma_{db}-i\Omega\sigma_{bd} =0\L {7} \\ &&\dot\sigma_{cc}
=\mu-\gamma_{0} \sigma_{cc} + \gamma\sigma_{aa} +
i\Omega^*\sigma_{ac}-i\Omega\sigma_{ca} =0\\ &&\dot\sigma_{dd}
=-\(2\gamma+\gamma_0\)\sigma_{dd} + i\Omega\sigma_{bd}-i\Omega^*\sigma_{db} =0\L
{2.6} \EY
and
  \BY
   &&\dot\sigma_{ab} = - \(\gamma +\gamma_0+i\Delta_p \)\sigma_{ab} +
i\Omega_p \(\sigma_{bb}-\sigma_{aa} \) + i\Omega\sigma_{cb} -i\Omega\sigma_{ad}
=0\L {9} \\
 &&\dot\sigma_{ca} = -
\(\gamma+\gamma_0-i\Delta_1 \)\sigma_{ca}- i\Omega_p^*\sigma_{cb}-i\Omega^*
\(\sigma_{cc}-\sigma_{aa} \) =0\L {10} \\
&&\dot\sigma_{ad} = - \(2\gamma +\gamma_0+i\Delta_p-i\Delta_2 \)\sigma_{ad} +
i\Omega_p\sigma_{bd} + i\Omega\sigma_{cd}-i\Omega^*\sigma_{ab} =0\L {11} \\
&&\dot\sigma_{cb} = - \(\gamma_0 +i\Delta_p-i\Delta_1 \)\sigma_{cb}-
i\Omega_p\sigma_{ca} + i\Omega^*\sigma_{ab} -i\Omega\sigma_{cd} =0\L {12} \\
&&\dot\sigma_{bd} = - \(\gamma+\gamma_0-i\Delta_2 \)\sigma_{bd} + i\Omega^*
\(\sigma_{dd}-\sigma_{bb} \) + i\Omega_p^*\sigma_{ad} =0\L {13} \\
&&\dot\sigma_{cd} = - \(\gamma +\gamma_0+i\Delta_p-i\Delta_1-i\Delta_2
\)\sigma_{cd} + i\Omega^*\sigma_{ad}-i\Omega^*\sigma_{cb} =0\L {}
 \EY
Under writing the equations the following replacements have been made:
 \BY
&&\sigma_{ab} \to\sigma_{ab} e^{-i \Delta_p t}
\\
 &&\sigma_{ac} \to\sigma_{ac} e^{-i \Delta_1 t} \\
&&\sigma_{db} \to\sigma_{db} e^{-i \Delta_2 t}
\\
 &&\sigma_{ad} \to\sigma_{ad} e^{-i \(\Delta_p-\Delta_2 \) t} \\
&&\sigma_{dc} \to\sigma_{dc} e^{-i \(\Delta_p-\Delta_1-\Delta_2 \) t}
\\
 &&\sigma_{bc} \to\sigma_{bc} e^{-i \(\Delta_p-\Delta_1 \) t}\L{}
  \EY
We put all the derivatives equaled to zero because further only the stationary
solutions are of our interest.

Now to make the mathematical situation simpler we remember interestingthat the
probe wave is weak. It means we want to be restricted by the approximation
$\Omega_p/(\gamma+\gamma_0)\ll1$ and to remain in our expressions only the main
non-zero terms. As a result the wished coherencies read:
\BY
&&\sigma_{ab} = i\alpha_p \;\frac{1 +i f}{D} \[n_{ba} +n_{ba} \;
\frac{\mid\alpha\mid^2}{1 +i f} \(\frac{1}{2 -\varepsilon_0+i f_{p2}} +
\frac{1}{\varepsilon_0 +i f_{p1}} \)-\right.\nonumber \\ &&\left.-\; n_{ca} \;
\frac{1}{1-i f_1} \frac{\mid\alpha\mid^2}{\varepsilon_0 + i f_{p1}} + \; n_{db}
\; \frac{1}{1-i f_2} \frac{\mid\alpha\mid^2}{2 -\varepsilon_0+i f_{p2}}
\], \L{2.19}\\
&&\sigma_{ca} = - n_{ca}\;\frac{i\alpha^*}{1-i f_1},\L{2.20} \\
&&\sigma_{bd} = - n_{bd}\;\frac{i\alpha^*}{1-i f_2}.\L{2.21}
\EY
The denominator $D$ in (\ref{2.19})  is given by:
\BY
 && D = \(1 +i f \) \(1 +i f_p \) +\mid\alpha\mid^2 \(2 +i
f_{p1} + i f_{p2} \) \(\frac{1}{\varepsilon_0 +i f_{p1}} + \frac{1}{2
-\varepsilon_0+i f_{p2}} \). \L{16}
\EY
Here are  the following notations:
\BY
 && n_{ik} = \sigma_{ii}-\sigma_{kk}, \qquad i, k=a, b, c, d \\
 &&f_p
=\frac{\Delta_p}{\gamma(1+\varepsilon_0)}, \qquad f_1 =
\frac{\Delta_1}{\gamma(1+\varepsilon_0)}, \qquad f_2
=\frac{\Delta_2}{\gamma(1+\varepsilon_0))},\qquad \varepsilon_0
=\frac{\gamma_0}{\gamma+\gamma_0}, \\
&&f_{p1} =f_p-f_1, \qquad f_{p2} =f_p-f_2, \qquad f=f_p-f_1-f_2, \qquad f_{12}
=f_1-f_2
\\
&&\alpha_p =\frac{\Omega_p}{\gamma(1+\varepsilon_0)}, \qquad \alpha
=\frac{\Omega}{\gamma(1+\varepsilon_0)}, \qquad \alpha_0 =\frac{\alpha}{\sqrt
{\varepsilon_0}}, \qquad n_0 = \frac{\mu}{\gamma_0}
 \EY
Substituting (\ref{2.19}-\ref{2.21}) into (\ref{2.3}-\ref{2.6}) we obtain the
closed system of equation for all the level populations. After relatively simple
operations we are able to get the populations in the explicit form:
 \BY
&&\sigma_{aa}=n_0\frac{2\mid\alpha\mid^2}{\(2-\varepsilon_0\)\(1+f_1^2\)
+2\mid\alpha_0\mid^2\(1+\varepsilon_0\)}\L{2.27}\\
&&\sigma_{bb}=n_0\frac{\(2-\varepsilon_0\)\(1+f_1^2\)+4\mid\alpha_0\mid^2}{\(2-\varepsilon_0\)
\(1+f_1^2\)+2\mid\alpha_0\mid^2\(1+\varepsilon_0\)}\;\frac{\(2-\varepsilon_0\)\(1+f_2^2\)+2\mid\alpha\mid^2}
{\(2-\varepsilon_0\) \(1+f_2^2\)+4\mid\alpha\mid^2}\\
&&\sigma_{cc}=n_0\frac{\(2-\varepsilon_0\)\(1+f_1^2\)+2\mid\alpha\mid^2}{\(2-\varepsilon_0\)
\(1+f_1^2\)+2\mid\alpha_0\mid^2\(1+\varepsilon_0\)}\\
&&\sigma_{dd}=n_0\frac{\(2-\varepsilon_0\)\(1+f_1^2\)+4\mid\alpha_0\mid^2}{\(2-\varepsilon_0\)
\(1+f_1^2\)+2\mid\alpha_0\mid^2\(1+\varepsilon_0\)}\;\frac{2\mid\alpha\mid^2}{\(2-\varepsilon_0\)
\(1+f_2^2\)+4\mid\alpha\mid^2}\L{2.30}
 \EY
Substituting these formulas into (\ref{2.19})-(\ref{2.21}) we can find all the
coherencies in the explicit form and in particular the  $\sigma_{ab}$ which is
important for analysis of the optical properties of the $(ab)$-transition.

  Further everywhere we will  be restricted by
the small value $\varepsilon_0\ll1$. It means our requirement to the system is
the passed lifetime of the atom through the light beam volume is much more than
the lifetime of the upper laser state connected with spontaneous decay.

\section{The thermal motion of the atoms}
\L{III} As mentioned above our main goal here is to discuss the situation with
the thermal motion of the atoms, forming  the output signal under crossing the
light beam volume. To describe correctly this case we have no right to use
directly the previous formulas, because they are suitable in the obtained form
only for the immovable atoms. But it is well known how to generalize it on the
case of the thermal ensemble. We need to introduce by hands the available
Doppler shifts, depending on velocities of the atoms,  and then to average the
formulas over the velocities  with the adequate velocity distribution. In our
conditions it means we have to make the following replacements in our formulas
(\ref{2.19}) and (\ref{2.27})-(\ref{2.30}):
 \BY
  && f_p\to f_p+x, \qquad f_1\to f_1+x, \qquad
f_2\to f_2+x,\qquad f\to f-x
\EY
and then to average the obtained expressions with the Maxwell's distribution:
\BE
 W (x) = \frac{1}{\sqrt {\pi} x_0} e^ {-x^2/x_0^2}, \qquad x_0
=\frac{ku}{\gamma},
 \EE
where $ku$ is the Doppler spectral width of the absorption contour. In our
approach we choose the so-called Doppler limit $x_0\gg1$.  One can find all the
details of these operations in Appendix \ref{App1}.

After that we have to make additionally two things. First, we want to be
restricted by case when the drive field frequency coincides with the atomic
transition frequency in the absence of the magnetic field. Second, we have to
introduce into the formulas the magnetic field in the explicit form. To satisfy
this program we must make the following replacements in our formulas:
\BE
 f_1\to
0, \qquad f_2\to-h, \qquad f_p\to f_p-h, \qquad f_{p1} \to f_p-h, \qquad f_{p2}
\to f_p, \qquad f\to f_p
\EE
where $h$ is the Zeeman shift of the $(b)$-level.

 These operations are carried
out in Appendix \ref{App2}.

\section {Atoms at rest}\L{IV}
In this section we will discuss the case of the immovable atoms. We think this
can be interesting and useful for understanding the situation as whole.

According to (\ref{2.19}) we can write the coherency for the $(ab)$-transition
as the sum of three terms:
\BY
&&\sigma_{ab} = \sigma_{ab}^{ab} + \sigma_{ab}^{ac} + \sigma_{ab}^{bd},\L{3.1}
\EY
connected respectively with the population differences $n_{ab},\;n_{ac}$ and
$n_{bd}$ and having the following explicit forms:
 \BY
 &&
 \sigma_{ab}^{ab} = i\alpha_pn_0 \;\frac{1 +i f_p-2i f_1}{D} \; \frac{1+f_1^2+2\mid\alpha_0\mid^2}{1+f_1^2 +\mid\alpha_0\mid^2}\;
\frac{1+\(f_1-i h\)^2+\mid\alpha\mid^2}{1+\(f_1-i h\)^2
+2\mid\alpha\mid^2}\times\nn\\ && \[1 +\frac{\mid\alpha\mid^2}{1 +i f_p-2i f_1}
\(\frac{1}{\varepsilon_0 +i f_{p1}} + \frac{1}{2 +i f_{p1}+i h} \)
\], \\&&\nonumber \\
&&\sigma_{ab}^ {ac} =-i\alpha_p\;n_0\frac{1 +i f_p-2i f_1}{D}\;\frac{1 +i
f_1}{1+f_1^2 +\mid\alpha_0\mid^2}\;\frac{\mid\alpha\mid^2}{\varepsilon_0 +i
f_{p1}},
\\&&\nonumber \\
&&\sigma_{ab}^{db} =- i\alpha_pn_0 \;\frac{1 +i f_p-2i f_1}{D} \;
\frac{1+f_1^2+2\mid\alpha_0\mid^2}{1+f_1^2 +\mid\alpha_0\mid^2}\; \frac{1+i
f_1-i h}{1+\(f_1-i h\)^2 +2\mid\alpha\mid^2}\times\nn\\
&& \frac{\mid\alpha\mid^2}{2 +i f_{p1}+i h},\\&&\nn\\
&&\mbox{where}\nn\\
&& D = \(1 +i f_p-2i f_1 \) \(1 +i f_p-ih \) +\mid\alpha\mid^2 \(2 +2i f_{p1} +
i h \)\times\nn\\
&& \(\frac{1}{\varepsilon_0 +i f_{p1}} + \frac{1}{2+i f_{p1}+i h} \)\L{3.5}
\EY
 These equations describe the absorption $Im(\sigma_{ab})$  and dispersion
 $Re(\sigma_{ab})$ of the probe wave
 interacting with $(ab)$-transition.

Here we have introduced the external magnetic field in the explicit form. We
have put in the absence of the magnetic field  the upper and lower  levels are
quite degenerated. It means the frequency shifts of the drive wave relative to
both
  the transitions $(ac)$ and $(bd)$ are the same, that is $f_1=f_2$.
  For simplicity we have assumed that in the magnetic field only the
$(b)$-level moves. Then  $f_2=f_1-h$, where $h$ - is the
Zeeman's shift of the $(b)$-level.

The formulas (\ref{3.1})-(\ref{3.5}) can be rewritten in the regime of the
saturation by the drive field $\mid\alpha\mid,\mid\alpha_0\mid\gg1$ in the
simpler and more visible form. In resonant drive tuning $f_1=0$ and with the
high power of the external magnetic field $(h\gg1)$
 the absorptive and dispersive contours read:
 \BY
 &&
Re\(\sigma_{ab}/\( \alpha_pn_0\)\)
=\frac{2\(f_p-h-\mid\alpha\mid\)}{1+4\(f_p-h-\mid\alpha\mid\)^2}
+\frac{2\(f_p-h+\mid\alpha\mid\)}{1+4\(f_p-h+\mid\alpha\mid\)^2}\\
&& Im\(\sigma_{ab}/\( \alpha_pn_0\)\)=
\frac{1}{1+4\(f_p-h-\mid\alpha\mid\)^2}+\frac{1}{1+4\(f_p-h+\mid\alpha\mid\)^2}
 \EY
One can see these results can be interpreted as the dynamical Stark effect for
three-level atom. Really, here the initial Lorenzian (or dispersive curve) with
the zero drive field, placed on $f_p=h$, is split in the non-zero drive field on
the two ones, placed on the horizontal axis symmetrically relative to $f_p=h$ on
the frequencies $h\pm|\alpha|$. This effect is very clear because under the
essential drift of the $(b)$-level  the $(d)$-level falls simply  out of the
interaction with the drive field and so effectively the four-level atom is
converted into the three-level one.

 To the contrary, without the  magnetic field $(h=0)$ all the four levels
  take an important role. The spectral contours in the case are given as:
 \BY
 &&Re\(\sigma_{ab}/\( \alpha_pn_0\)\)=\frac{f_p}{1+f_p^2}+\frac{1}{2}\;
\frac{f_p-2\mid\alpha\mid}{1+\(f_p-4\mid\alpha\mid\)^2}+\frac{1}{2}\;\frac{f_p+2\mid\alpha\mid}
{1+\(f_p+4\mid\alpha\mid\)^2}\L{4.8}\\
&&Im(\sigma_{ab}/(\alpha_pn_0))=\frac{1}{1+f_p^2}+\frac{1}{2}\;\frac{1}{1+\(f_p-4\mid\alpha\mid\)^2}+
\frac{1}{2}\;\frac{1}{1+\(f_p+4\mid\alpha\mid\)^2}\L{4.9}
\EY
Now  there are three Lorencians in (\ref{4.8}) (or three dispersive curves in
(\ref{4.9})) instead of two in the previous situation. This difference is easy
understood on the qualitative level under applying the model of the
Rabi-splitting of levels. Without the magnetic field (the N-configuration for
the four-level atom) both levels $(a)$ and $(b)$ are split, that gives three
spectral contours instead of one. At the same time, in the strong magnetic field
only the $(a)$-level is split, what ensures only two contours. In Fig.\ 3 the
formulas (\ref{4.8})-(\ref{4.9}) are presented graphically with
$\mid\alpha\mid^2=10$.

At the same time, as stated above in the non-zero magnetic field we must expect
a gradual transformation of the four-level atom into the three-level one. In
Fig.\ 4 the curves are drown with
$h=7,\;\varepsilon_0=1/10,\;\mid\alpha\mid^2=10$. One can see already with not
very strong magnetic field the curves in the main match to the three-level
structure in $\Lambda$-configuration.

The EIT effect is available as we remember for the three-level atom. In
consequence of the Rabi splitting in the strong drive field the center of the
spectral contour turns out just  between two peaks that is in the transparency
area. Because in our case under gradual raising the magnetic field  the
three-level atom is achieved we can expect the stronger magnetic field, the
better EIT. This is demonstrated in Fig.\ 5. There we watch over the center   of
the spectral contour $f_p=h$ and show, how this point falls in relation to $h$.

One can see the EIT effect takes a place even in the zero magnetic field (h=0)
and the effect raises with $h$ as it was expected. The EIT effect with $h=0$
corresponds to falling the central point of the absorption contour in the strong
drive field (see Fig.\ 3).

\section{The non-linear interference effects in the N-configuration\L{V}}

\subsection{The EIA, EIT effects in the  atomic vapor}
Now let us discuss the experimental situation with the thermal ensemble of the
atoms. In Sec. \ref{III} it has been described, how to get the available
formulas for this case. In App. \ref{App1} and \ref{App2} this program has been
implemented.

In this section we are going to discuss some interesting details in the
behaviour of the coherency $\sigma_{ab}$ in relation to the drive field and the
externally applied magnetic field. Our  consideration will be based on numerical
calculations by the formulas (\ref{B.2})-(\ref{B.11}), which have been given for
the case of exact frequency tuning the drive field ($f_1=0$).

First of all, let us discuss the absorptive properties of the vapor relative to
the probe field. This is determined by  the imaginary part of the coherency
$\sigma_{ab}$, and this is presented in  Fig.~6a for the zero magnetic field
$h=0$ and with $x_0=100, \; \varepsilon_0=1/10, \; \mid\alpha_0\mid=10$.

Everywhere further we discuss not  the density matrix itself but the value
$\sigma/\lambda$. The factor $\lambda  $ is chosen so to normalize this value
onto one for the zero drive field in the contour center. So one can conclude the
absorption with the strong drive field in the point $f_p=0$ is much more (with
factor about 5) than without the drive field. It means in the zero magnetic
field the N-configuration ensures the essential EIA effect.

At the same time for the high enogh magnetic field in the point $f_p=h$ already
the EIT effect takes a place. It is demonstrated in Fig.\ 7a  for case $h=10 $.
One can see in the point $f_p=10$ the absorption is much less than one (about
0,165).

As a result we can conclude in the N-configuration we have a possibility to
control an interference changing it by the externally applied magnetic field
from distructive to constructive and back.

\subsection{The slow light effect in the  atomic vapor}
 Let us discuss now dispersive properties of the vapor on the
$(ab)$-transition. For that we need to investigate the real part of the matrix
element $\sigma_{ab}$.  The respective frequency dependences are presented in
Fig.\ 6b for the zero magnetic field $(h=0)$ and in Fig.\ 7b for $h=10$.

The interesting areas connect with a big derivatives with respect to
$f_p(\omega_p)$. Just there we can expect essential slowing  the light pulse
down. In the zero magnetic field (Fig.\ 6b) this is in the area near the zero
frequency. At the same time with $h=10$ this is in the area near $f_p=10$
$(f_p=h)$.
 But the  effect in the zero magnetic field (on the zero frequency)
 is bad for observation, because
here the very effective absorption takes a place (compare with Fig.\ 6a).

At the same time, in the magnetic field (Fig.\ 7b) in the vicinity of $f_p=10$,
where the good EIT effect takes a place, the derivative with respect to $f_p$ is
high enough too. Taking this into account further we will analyze the group
velocity of the light pulse as the function of the magnetic field in this area.

As is known the group velocity $v_{gr.}$ can be expressed via a real part of the
susceptibility $\chi=Re\chi+iIm\chi$ as:
 \BY
&&\frac{v_{gr.}}{c}=\[1+2\pi\(1+\omega_p\frac{d}{d\omega_p}\)Re\chi\]^{-1}
\EY
 Rewriting it in terms of the density matrix and
in our notations the formula reads:
 \BY
 &&\frac{v_{gr.}}{c}=\[1+{\cal
A}\(1-\frac{\omega_p}{\gamma}\frac{d}{df_p}\)Re\sigma_{ab}\]^{-1},\qquad{\cal
A}=\frac{3}{16\pi^2}\lambda_p^{3}N\frac{\gamma_{rad.}}{\gamma}\frac{1}{\mid\alpha\mid}
\EY
 Here $\lambda_p$ and $\omega_p$ are the wavelength and the frequency of the probe
field, $N$ is the atomic concentration, $\gamma_{rad.}$ is the constant of the
radiation decay. For a numerical calculation  we choose the following set of the
parameters: $ \lambda_p\sim~10^{-4}cm,\; N\sim~10^{12}cm^{-3},
\;\gamma_{rad.}/\gamma\sim~1,
\;\omega_p/\gamma\sim~{10}^8,\;|\alpha_p/\alpha|\sim10^{-1}$.

In Fig.\ 8 the dependence of the group velocity in the vicinity of $f_p =h$ (the
EIA or EIT area)  on the magnetic field $h$ is presented with $|\alpha_0|=10$.
As was expected the minimum of the velocity  is achieved in the zero magnetic
field. But we remember the light here is strongly absorbed on the transition. At
the same time, with $h>5$ there is a slow light effect too, and there
$v_{gr.}\sim 10^{-4}c$. This depends actually on the power of the drive field.
One can see in Fig.\ 9 with $f_p=h=10$ at first the group velocity increases
with changing the amplitude $\alpha_0$ from small to higher meanings, passes the
maximum and next falls.

In conclusion, we have proposed effective
coherent control of the optical properties
of the resonant thermal medium. Varying  the external magnetic field we can
achieve in the same experiment  both the EIT and EIA effects and also slow down
light or to the contrary accelerate.

\begin{acknowledgments}
This work was performed within the Franco-Russian cooperation program ``Lasers
and Advanced Optical Information Technologies'' with financial support from the
following organizations: INTAS (grant INTAS-01-2097), RFBR (grant 03-02-16035),
Minvuz of Russia (grant E 02-3.2-239), and by the Russian program
``Universities of Russia'' (grant ur.01.01.041).
Also we gratefully
acknowledge the support from the Office of Naval Research, the
Air Force Research Laboratory (Rome, NY), Defense Advanced
Research Projects Agency-QuIST, Texas A$\&$M University
Telecommunication and Information Task Force (TITF) Initiative,
and the Robert A. Welch Foundation.

\end{acknowledgments}

\appendix

\section {Thermal motion of atoms}
\L{App1}
 To take into account thermal motion of atoms   we should  make the
following frequency replacements  in our formulas:
\BY
&& f_p\to f_p+x, \qquad f_1\to
f_1+x, \qquad f_2\to f_2+x,\qquad f=\to f-x  \\
&&( f_{p1} =f_p-f_1\to f_{p1},
\qquad f_{p2} =f_p-f_2\to f_{p2}, \qquad f_{12} =f_1-f_2\to f_{12})\nn
 \EY
  and
then to average the formulas over $x$ with the Maxwell's distribution:
 \BE
 W (x) = \frac{1}{\sqrt {\pi}
x_0} e^ {-x^2/x_0^2}, \qquad x_0 =\frac{ku}{\gamma}
\EE
After  these operations in the Doppler limit ($x_0\gg1 $) the coherency,
normalized on unity in the center of line of absorption with the zero driving
field $\mid\alpha\mid$, is represented as the sum of terms:
\BE
\sigma_{ab}/\lambda=\sigma_{ab}^{ab}/\lambda+\sigma_{ab}^{ac}/\lambda
+\sigma_{ab}^{bd}/\lambda, \qquad\lambda =\alpha_p n_0\sqrt{\pi}/x_0
 \EE
  where
   \BY
   &&\sigma_{ab}^{ab}/\lambda=A_1+A_2+A_3+A_4+A_5\\&&\nn \\
   && A_1 =\frac{1}{\pi}
\int_{-\infty}^ {+ \infty} e^ {-x^2/x_0^2} \frac{1}{x+f_p-i} \; \frac{1 +
\(f_1+x \)^ 2+2\mid\alpha_0\mid^2}{1 + \(f_1+x \)^ 2 +\mid\alpha_0\mid^2} \;
dx=A_1^\prime +i A_1^ {\prime\prime} \\
&& A_2 =-\;\mid\alpha\mid^2 \(2 +i f_{p1} + i f_{p2} \) \(\frac{1}{\varepsilon_0
+i f_{p1}} + \frac{1}{2 +i f_{p2}} \)\frac{1}{\pi} e^ {-f_p^2/x_0^2}
\times\nonumber \\ && \int_{-\infty}^ {+ \infty} \frac{1}{x+f_p-i} \; \frac{1 +
\(f_1+x \)^ 2+2\mid\alpha_0\mid^2}{1 + \(f_1+x \)^ 2 +\mid\alpha_0\mid^2} \;
\frac{1}{\(x-x_1 \) \(x-x_2 \)} \; dx \\
&&A_3=i\mid\alpha\mid^2\(\frac{1}{\varepsilon_0+i f_{p1}}+\frac{1}{2+i
f_{p2}}\)\frac{1}{\pi} e^{-f_p^2/x_0^2}\times\nn\\
&& \int\limits_{-\infty}^{+\infty}
\frac{1+\(f_1+x\)^2+2\mid\alpha_0\mid^2}{1+\(f_1+x\)^2+\mid\alpha_0\mid^2}\;
\frac{1}{\(x-x_1\)\(x-x_2\)}dx\\
&& A_4 =-i\mid\alpha\mid^2\frac{1}{\pi} e^ {-f_p^2/x_0^2} \times\nonumber \\
&& \int_{-\infty}^ {+ \infty}
 \frac{1 + \(f_1+x \)^ 2+2\mid\alpha_0\mid^2}{1 + \(f_1+x \)^ 2 +\mid\alpha_0\mid^2} \;
\frac{1+i\(f-x\)}{\(x-x_1 \) \(x-x_2
\)}\frac{1}{1+\(f_2+x\)^2+2\mid\alpha\mid^2} dx\\&&\nn\\
&& A_5 =-i\mid\alpha\mid^4 \(\frac{1}{\varepsilon_0 +i f_{p1}} + \frac{1}{2 +i
f_{p2}}\) \frac{1}{\pi} e^ {-f_p^2/x_0^2} \times\nonumber \\
&& \int_{-\infty}^ {+ \infty}
 \frac{1 + \(f_1+x \)^ 2+2\mid\alpha_0\mid^2}{1 + \(f_1+x \)^ 2 +\mid\alpha_0\mid^2} \;
\frac{1}{\(x-x_1 \) \(x-x_2 \)}\frac{1}{1+\(f_2+x\)^2+2\mid\alpha\mid^2}
dx\\
&&\nn\\
&&\sigma_{ab}^ {ac}/\lambda =-\;\frac{\mid\alpha\mid^2}{\varepsilon_0 +i f_{p1}}
\; \frac{1}{\pi} e^ {-f_p^2/x_0^2}  \times\nn\\
&& \int_{-\infty}^ {+ \infty} \frac{x-f +i}{\(x-x_1 \) \(x-x_2 \)}\frac{1 +i
\(f_1+x \)}{1 + \(f_1+x \)^ 2 +\mid\alpha_0\mid^2} \; \; dx
\\
&&\nonumber \\
&&\sigma_{ab}^ {bd}/\lambda =-i\;\frac{\mid\alpha\mid^2}{2 +i f_{p2}} \;
\frac{1}{\pi} e^ {-f_p^2/x_0^2}\times\nn\\ && \int\limits_{-\infty}^ {+
\infty}\frac{1}{\(x-x_1 \) \(x-x_2 \)}\; \frac{\(1-i x+i f\)\(1+i f_2+i
x\)}{1+\(f_2+x\)^2+2\mid\alpha\mid^2} \; \frac{1 +
 \(f_1+x \)^ 2+2\mid\alpha_0\mid^2}{1 + \(f_1+x \)^ 2 +\mid\alpha_0\mid^2} dx
\EY
Here
   \BY
   &&x_{1,2} = - \;\frac{1}{2} \(f_1+f_2 \)\pm\frac{i}{2} \(2 +i
f_{p1} + i f_{p2} \) \sqrt {1 + \frac{4\mid\alpha\mid^2}{\(\varepsilon_0 + i
f_{p1} \) \(2 +i f_{p2} \)}}
 \EY
  are the roots of the quadratic relative to $x$
equation
\BY
 && \(1 +i f-i x \) \(1 +i f_p +i x \) +\mid\alpha\mid^2 \(2 +i
f_{p1} + i f_{p2} \) \(\frac{1}{\varepsilon_0 +i f_{p1}} + \frac{1}{2 +i f_{p2}}
\) =0
 \EY
  Now we are able  to carry out all integrations of the values of the
complex resudues and obtain the following formulas expressed via the physical
parameters  in the explicit forms :
 \BY &&\sigma_{ab}^
{ab}/\lambda=A_1+A_2+A_3+A_4+A_5, \L{A14}\\
&& Re\(A_1\) =\frac{1}{\pi} \int_{-\infty}^ {+ \infty} e^ {-x^2/x_0^2}
\frac{x+f_p}{\(x+f_p \)^ 2+1} \; \frac{1 + \(f_1+x \)^ 2+2\mid\alpha_0\mid^2}{1
+ \(f_1+x \)^ 2 +\mid\alpha_0\mid^2} \; dx \\
 && Im\(A_1\) =e^ {-f_p^2/x_0^2}\times\nn\\
 &&
\[\frac{f_{p1} \(f_{p1}-2i \) + 2\mid\alpha_0\mid^2}{f_{p1} \(f_{p1}-2i \)
+\mid\alpha_0\mid^2} + \frac{\mid\alpha_0\mid^2}{\sqrt {1 +\mid\alpha_0\mid^2}
\[f_{p1} \(f_{p1} +2i\sqrt{1 +\mid\alpha_0\mid^2} \)-\mid\alpha_0\mid^2 \]} \]
\\ && A_2 =\mid\alpha\mid^2 \(2 +i f_{p1} + i f_{p2} \)
\(\frac{1}{\varepsilon_0 +i f_{p1}} + \frac{1}{2 +i f_{p2}} \) e^ {-f_p^2/x_0^2}
\times\nonumber \\ && \[\frac{\mid\alpha_0\mid^2}{\sqrt {1 +\mid\alpha_0\mid^2}
\(i\sqrt {1 +\mid\alpha_0\mid^2}-f_{p1} + i \) \(i\sqrt {1 +\mid\alpha_0\mid^2}
+f_1+x_1 \) \(i\sqrt {1 +\mid\alpha_0\mid^2} +f_1+x_2 \)} +\right.\nonumber
\\&&\nonumber \\ &&\left. + \;\frac{1 + \(f_1+x_2 \)^ 2+2\mid\alpha_0\mid^2}{1 +
\(f_1+x_2 \)^ 2 +\mid\alpha_0\mid^2} \; \frac{2i}{\(x_2+f_p-i \) \(x_2-x_1 \)}
\] \\ && A_3 =-2\mid\alpha\mid^2 \(\frac{1}{\varepsilon_0 +i f_{p1}} +
\frac{1}{2 +i f_{p2}} \) e^ {-f_p^2/x_0^2} \[\frac{1 + \(f_1+x_1 \)^
2+2\mid\alpha_0\mid^2}{1 + \(f_1+x_1 \)^ 2 +\mid\alpha_0\mid^2} \;
\frac{1}{\(x_1-x_2 \)} +\right.\nonumber \\ &&\left.+ \;
\frac{\mid\alpha_0\mid^2}{2i\sqrt{1 +\mid\alpha_0\mid^2} \(i\sqrt{1
+\mid\alpha_0\mid^2}-f_1-x_1 \) \(i\sqrt{1 +\mid\alpha_0\mid^2}-f_1-x_2\)}\]
\\
&& A_4 =2\mid\alpha\mid^2 e^ {-f_p^2/x_0^2}\[\frac{1 + \(f_1+x_1
\)^2+2\mid\alpha_0\mid^2} {1 + \(f_1+x_1 \)^2 +\mid\alpha_0\mid^2} \; \frac{1+i
f-i x_1}{x_1-x_2 }\; \frac{1}{1 + \(f_1+x_2 \)^2 +2\mid\alpha\mid^2}
+\right.\nonumber \\ &&\left. + \frac{\mid\alpha_0\mid^2\(1+i
f_{p2}+\sqrt{1+\mid\alpha_0\mid^2}\)} {2i\sqrt {1 +\mid\alpha_0\mid^2}\(i\sqrt
{1 +\mid\alpha_0\mid^2}-f_1-x_1 \) \(i\sqrt {1 +\mid\alpha_0\mid^2}-f_1-x_2
\)}\times\right.\nn\\
&&\left.\frac{1}{f_{21}\(f_{21}+2i\sqrt{1+\mid\alpha_0\mid^2}\)-\mid\alpha_0\mid^2}+\right.\nn\\
&&\left. +  \frac{1+i f_{p1}+\sqrt{1+2\mid\alpha\mid^2} }{2i\sqrt{1
+2\mid\alpha\mid^2} \(i\sqrt{1 +2\mid\alpha\mid^2}+f_1+x_1 \) \(i\sqrt{1
+2\mid\alpha\mid^2}+f_1+x_2 \) }\times\right.\nn\\
&&\left.\frac{f_{21}\(f_{21}-2i\sqrt{1+2\mid\alpha\mid^2}\)
+2\mid\alpha_0\mid^2}{f_{21}\(f_{21}-2i\sqrt{1+2\mid\alpha\mid^2}\)+\mid\alpha_0\mid^2}\]\\
&& A_5 =2\mid\alpha\mid^4 \(\frac{1}{\varepsilon_0 +i f_{p1}} + \frac{1}{2 +i
f_{p2}} \) e^ {-f_p^2/x_0^2} \nn\\&&
 \[\frac{1 + \(f_1+x_1 \)^2+2\mid\alpha_0\mid^2}
{1 + \(f_1+x_1 \)^2 +\mid\alpha_0\mid^2} \; \frac{1}{x_1-x_2 }\; \frac{1}{1 +
\(f_1+x_2 \)^2 +2\mid\alpha\mid^2} +\right.\nonumber \\ &&\left. + \;
\frac{\mid\alpha_0\mid^2}{2i\sqrt {1 +\mid\alpha_0\mid^2} \(i\sqrt {1
+\mid\alpha_0\mid^2}-f_1-x_1 \) \(i\sqrt {1 +\mid\alpha_0\mid^2}-f_1-x_2 \)
}\times\right.\nn\\
&&\left.\frac{1}{f_{21}\(f_{21}+2i\sqrt{1+\mid\alpha_0\mid^2}\)-\mid\alpha_0\mid^2}+\right.\nn\\
&&\left. + \; \frac{1}{2i\sqrt{1 +2\mid\alpha\mid^2} \(i\sqrt{1
+2\mid\alpha\mid^2}+f_1+x_1 \) \(i\sqrt{1 +2\mid\alpha\mid^2}+f_1+x_2
\)}\times\right.\nn\\ &&\left.\frac{
f_{21}\(f_{21}-2i\sqrt{1+2\mid\alpha\mid^2}\) +2\mid\alpha_0\mid^2}{
f_{21}\(f_{21}-2i\sqrt{1+2\mid\alpha\mid^2}\)+\mid\alpha_0\mid^2}\]
\\
&&\nn\\ &&\nn\\ &&\sigma_{ab}^ {ac}/\lambda
=\frac{2\mid\alpha\mid^2}{\varepsilon_0 +i f_{p1}} e^ {-f_p^2/x_0^2}
\[\frac{\(1-\sqrt {1 +\mid\alpha_0\mid^2} \) \(f_{p2}-i-i\sqrt {1
+\mid\alpha_0\mid^2} \)}{2\sqrt {1 +\mid\alpha_0\mid^2} \(i\sqrt {1
+\mid\alpha_0\mid^2}-f_1-x_1 \) \(i\sqrt {1 +\mid\alpha_0\mid^2}-f_1-x_2 \)}
+\right.\nonumber \\ &&\left. +\frac{x_1-f +i}{x_1-x_2} \; \frac{x_1+f_1-i}{1 +
\(f_1+x_1 \)^ 2 +\mid\alpha_0\mid^2} \] \\ &&\nonumber \\ &&\nonumber \\
&&\sigma_{ab}^ {bd}/\lambda =\frac{2\mid\alpha\mid^2}{2 +i f_{p2}} \; e^
{-f_p^2/x_0^2} \[\frac{1 + \(f_1+x_1 \)^2+2\mid\alpha_0\mid^2} {1 + \(f_1+x_1
\)^2 +\mid\alpha_0\mid^2} \; \frac{1+i f-i x_1}{\(x_1-x_2 \)}\; \frac{1-i f_1-i
x_2}{1 + \(f_1+x_2 \)^2 +2\mid\alpha\mid^2} +\right.\nonumber \\ &&\left. +
\frac{\mid\alpha_0\mid^2}{2i\sqrt {1 +\mid\alpha_0\mid^2} \(i\sqrt {1
+\mid\alpha_0\mid^2}-f_1-x_1 \) \(i\sqrt {1 +\mid\alpha_0\mid^2}-f_1-x_2
\)}\times\right.\nn\\ &&\left.\frac{\(1+i f_{p2}+\sqrt{1+\mid\alpha_0\mid^2}\)
\(1+i f_{21}-\sqrt{1+\mid\alpha_0\mid^2}\)}{
f_{21}\(f_{21}+2i\sqrt{1+\mid\alpha_0\mid^2}\)-\mid\alpha_0\mid^2}+\right.\nn\\
&&\left. + \frac{\(1-\sqrt{1 +2\mid\alpha\mid^2}\)\(1+i f_{p1}+\sqrt{1
+2\mid\alpha\mid^2}\)} {2i\sqrt{1 +2\mid\alpha\mid^2} \(i\sqrt{1
+2\mid\alpha\mid^2}+f_1+x_1 \) \(i\sqrt{1 +2\mid\alpha\mid^2}+f_1+x_2
\)}\times\right.\nn\\
&&\left.\frac{f_{21}\(f_{21}-2i\sqrt{1+2\mid\alpha\mid^2}\)
+2\mid\alpha_0\mid^2}{
f_{21}\(f_{21}-2i\sqrt{1+2\mid\alpha\mid^2}\)+\mid\alpha_0\mid^2}\]\L{A22}\EY

\section{}
\L{App2}

As is seen here all the frequency detunings survive but in our discussion in the
main sections of the article we take into account a simpler situation when the
driving field is in resonance with the transition $(ac)$ what means $f_1=0$, and
only one level $(b)$ moves in the magnetic field. Then to take into account the
magnetic Zeeman effect in the explicit form let us make the following frequency
transformations: \BE
 f_1\to
0, \qquad f_2\to-h, \qquad f_p\to f_p-h, \qquad f_{p1} \to f_p-h, \qquad f_{p2}
\to f_p, \qquad f\to f_p \EE
Here the value $h$ is the Zeeman's shift of the $(b)$ level in the dimensionless
units.

As a result we have basic formulas for our discussion in the form: \BY
&&\sigma_{ab}^ {ab}/\lambda=A_1+A_2+A_3+a_4+A_5, \L{B.2} \\&&\nonumber \\ &&
Re\(A_1\) =\frac{1}{\pi} \int_{-\infty}^ {+ \infty} e^ {-x^2/x_0^2}
\frac{x+f_p-h}{\(x+f_p-h \)^ 2+1} \; \frac{1+x^2+2\mid\alpha_0\mid^2}{1+x^2
+\mid\alpha_0\mid^2} \; dx \\ && Im\(A_1\) =e^ {-\(f_p-h \)^ 2/x_0^2}
\[\frac{\mid\alpha_0\mid^2}{\sqrt {1 +\mid\alpha_0\mid^2}
\[\(f_p-h \) \(f_p-h+2i\sqrt {1 +\mid\alpha_0\mid^2} \)-\mid\alpha_0\mid^2 \]} +\right.\nonumber \\
&&\nonumber \\&&\left. +\frac{\(f_{p}-h \) \(f_{p}-h-2i \) +
2\mid\alpha_0\mid^2}{\(f_{p}-h \) \(f_{p}-h-2i \) +\mid\alpha_0\mid^2} \] \\ &&
A_2 =\mid\alpha\mid^2 \(2+2i f_{p}-i h\) \(\frac{1}{\varepsilon_0 +i f_{p}-i h}
+ \frac{1}{2 +i f_{p} } \) e^ {-\(f_p-h \)^ 2/x_0^2} \times\nonumber \\&&
\[\frac{\mid\alpha_0\mid^2} {\sqrt {1 +\mid\alpha_0\mid^2} \(i\sqrt {1
+\mid\alpha_0\mid^2}-f_{p} +h +i \) \(i\sqrt {1 +\mid\alpha_0\mid^2} +x_1 \)
\(i\sqrt {1 +\mid\alpha_0\mid^2} +x_2 \)} +\right.\nonumber \\ &&\left. +
\;\frac{1+x_2^2+2\mid\alpha_0\mid^2}{1+x_2^2 +\mid\alpha_0\mid^2} \;
\frac{2i}{\(-x_1+f_p-i \) \(x_2-x_1 \)} \] \\ && A_3 =-2\mid\alpha\mid^2
\(\frac{1}{\varepsilon_0 +i f_{p}-i h} + \frac{1}{2 +i f_{p}} \) e^ {-\(f_p-h
\)^ 2/x_0^2} \[\frac{1+x_1^2+2\mid\alpha_0\mid^2}{1+x_1^2 +\mid\alpha_0\mid^2}
\; \frac{1}{\(x_1-x_2 \)} +\right.\nonumber \\ &&\left. + \;
\frac{\mid\alpha_0\mid^2}{2i\sqrt {1 +\mid\alpha_0\mid^2} \(i\sqrt {1
+\mid\alpha_0\mid^2}-x_1 \) \(i\sqrt {1 +\mid\alpha_0\mid^2}-x_2 \)} \]\\ && A_4
=2\mid\alpha\mid^2 e^ {-\(f_p-h\)^2/x_0^2}\[\frac{1 + x_1^2+2\mid\alpha_0\mid^2}
{1 +x_1^2 +\mid\alpha_0\mid^2} \; \frac{1+i f_p-i x_1}{x_1-x_2 }\; \frac{1}{1 +
x_2^2 +2\mid\alpha\mid^2} -\right.\nonumber \\ &&\left. -
\frac{\mid\alpha_0\mid^2\(1+i f_{p}+\sqrt{1+\mid\alpha_0\mid^2}\)} {2i\sqrt {1
+\mid\alpha_0\mid^2}\(i\sqrt {1 +\mid\alpha_0\mid^2}-x_1 \) \(i\sqrt {1
+\mid\alpha_0\mid^2}-x_2 \)}\times\right.\nn\\
&&\left.\frac{1}{h\(-h+2i\sqrt{1+\mid\alpha_0\mid^2}\)+\mid\alpha_0\mid^2}+\right.\nn\\
&&\left. +  \frac{1+i f_{p}-i h+\sqrt{1+2\mid\alpha\mid^2} }{2i\sqrt{1
+2\mid\alpha\mid^2} \(i\sqrt{1 +2\mid\alpha\mid^2}+x_1 \) \(i\sqrt{1
+2\mid\alpha\mid^2}+x_2 \) }\times\right.\nn\\
&&\left.\frac{h\(h+2i\sqrt{1+2\mid\alpha\mid^2}\)
+2\mid\alpha_0\mid^2}{h\(h+2i\sqrt{1+2\mid\alpha\mid^2}\)+\mid\alpha_0\mid^2}\]\\
&& A_5 =2\mid\alpha\mid^4 \(\frac{1}{\varepsilon_0 +i f_{p}-i h} + \frac{1}{2 +i
f_{p}} \) e^ {-\(f_p-h\)^2/x_0^2} \nn\\&&
 \[\frac{1 + x_1^2+2\mid\alpha_0\mid^2}
{1 + x_1^2 +\mid\alpha_0\mid^2} \; \frac{1}{x_1-x_2 }\; \frac{1}{1 + x_2 ^2
+2\mid\alpha\mid^2} -\right.\nonumber \\ &&\left. -\;
\frac{\mid\alpha_0\mid^2}{2i\sqrt {1 +\mid\alpha_0\mid^2} \(i\sqrt {1
+\mid\alpha_0\mid^2}-x_1 \) \(i\sqrt {1 +\mid\alpha_0\mid^2}-x_2 \)
}\times\right.\nn\\
&&\left.\frac{1}{h\(-h+2i\sqrt{1+\mid\alpha_0\mid^2}\)+\mid\alpha_0\mid^2}+\right.\nn\\
&&\left. + \; \frac{1}{2i\sqrt{1 +2\mid\alpha\mid^2} \(i\sqrt{1
+2\mid\alpha\mid^2}+x_1 \) \(i\sqrt{1 +2\mid\alpha\mid^2}+x_2
\)}\times\right.\nn\\ &&\left.\frac{ -h\(-h-2i\sqrt{1+2\mid\alpha\mid^2}\)
+2\mid\alpha_0\mid^2}{
-h\(-h-2i\sqrt{1+2\mid\alpha\mid^2}\)+\mid\alpha_0\mid^2}\] \EY \BY
&&\sigma_{ab}^ {ac}/\lambda =\frac{2\mid\alpha\mid^2}{\varepsilon_0 +i f_{p}-i
h} e^ {-\(f_p-h \)^ 2/x_0^2} \times \\ && \[\frac{\(1-\sqrt {1
+\mid\alpha_0\mid^2} \) \(f_{p}-i-i\sqrt {1 +\mid\alpha_0\mid^2} \)} {2\sqrt {1
+\mid\alpha_0\mid^2} \(i\sqrt {1 +\mid\alpha_0\mid^2}-x_1 \) \(i\sqrt {1
+\mid\alpha_0\mid^2}-x_2 \)} +\frac{x_1-f_p +i}{x_1-x_2} \; \frac{x_1-i}{1+x_1^2
+\mid\alpha_0\mid^2} \] \nn \\ &&\nonumber \\ &&\sigma_{ab}^ {bd}/\lambda
=\frac{2\mid\alpha\mid^2}{2 +i f_{p}} \; e^ {-\(f_p-h\)^2/x_0^2} \[\frac{1 +
x_1^2+2\mid\alpha_0\mid^2} {1 +x_1^2 +\mid\alpha_0\mid^2} \; \frac{1+i f_p-i
x_1}{x_1-x_2 }\; \frac{1-i x_2}{1 + x_2 ^2 +2\mid\alpha\mid^2} -\right.\nonumber
\\ &&\left. -\frac{\mid\alpha_0\mid^2}{2i\sqrt{1 +\mid\alpha_0\mid^2}
\(i\sqrt{1 +\mid\alpha_0\mid^2}-x_1 \) \(i\sqrt{1 +\mid\alpha_0\mid^2}-x_2
\)}\times\right.\nn\\ &&\left.\frac{\(1+i
f_{p}+\sqrt{1+\mid\alpha_0\mid^2}\)\(1-i h-\sqrt{1+\mid\alpha_0\mid^2}\)}{
h\(-h+2i\sqrt{1+\mid\alpha_0\mid^2}\)+\mid\alpha_0\mid^2}+\right.\nn\\ &&\left.
+ \frac{\(1-\sqrt{1 +2\mid\alpha\mid^2}\)\(1+i f_{p}-i h+\sqrt{1
+2\mid\alpha\mid^2}\)} {2i\sqrt{1 +2\mid\alpha\mid^2} \(i\sqrt{1
+2\mid\alpha\mid^2}+x_1 \) \(i\sqrt{1 +2\mid\alpha\mid^2}+x_2
\)}\times\right.\nn\\ &&\left.\frac{h\(h+2i\sqrt{1+2\mid\alpha\mid^2}\)
+2\mid\alpha_0\mid^2}{ h\(h+2i\sqrt{1+2\mid\alpha\mid^2}\)+\mid\alpha_0\mid^2}\]
\EY

\BY &&x_{1,2} =\frac{1}{2}h\pm\frac{i}{2} \(2 +2i f_{p}-i h\) \sqrt {1 +
\frac{4\mid\alpha\mid^2}{\(\varepsilon_0 + i f_{p}-i h \) \(2 +i f_{p} \)}}
\L{B.11} \EY

 \newpage

 \begin{figure}[t]
  \includegraphics[width=120mm]{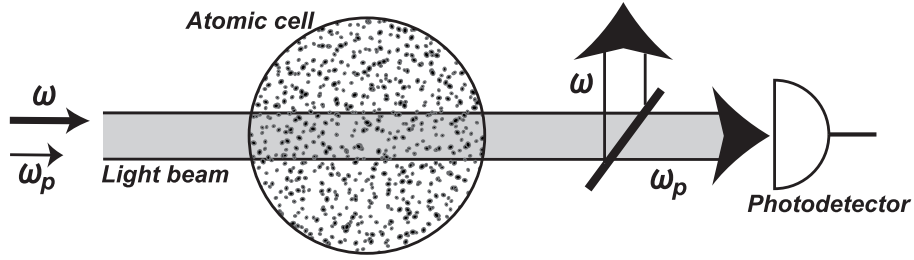}
 \caption{Experimental setup for observation of the EIT, EIA and slow light
 velocity effects}
 \label{fig1}
 \end{figure}

 \begin{figure}[t]
  \includegraphics[width=120mm]{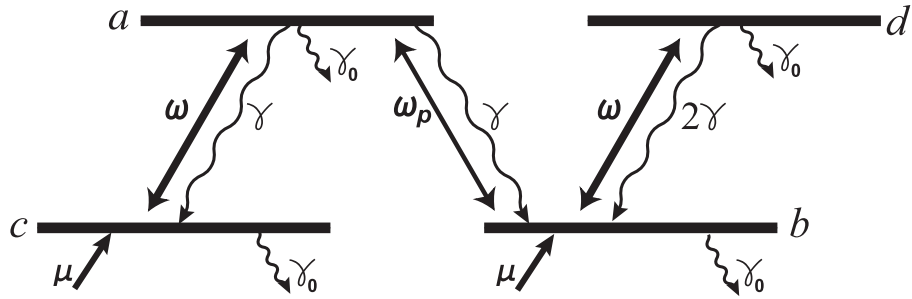}
 \caption{Schematic of N-type four-level system. Drive laser couples in the $(ac)$
 and $(bd)$ transitions simultaneously, probe laser to $(ab)$ }
 \label{fig2}
 \end{figure}

 \begin{figure}[t]
  \includegraphics[width=120mm]{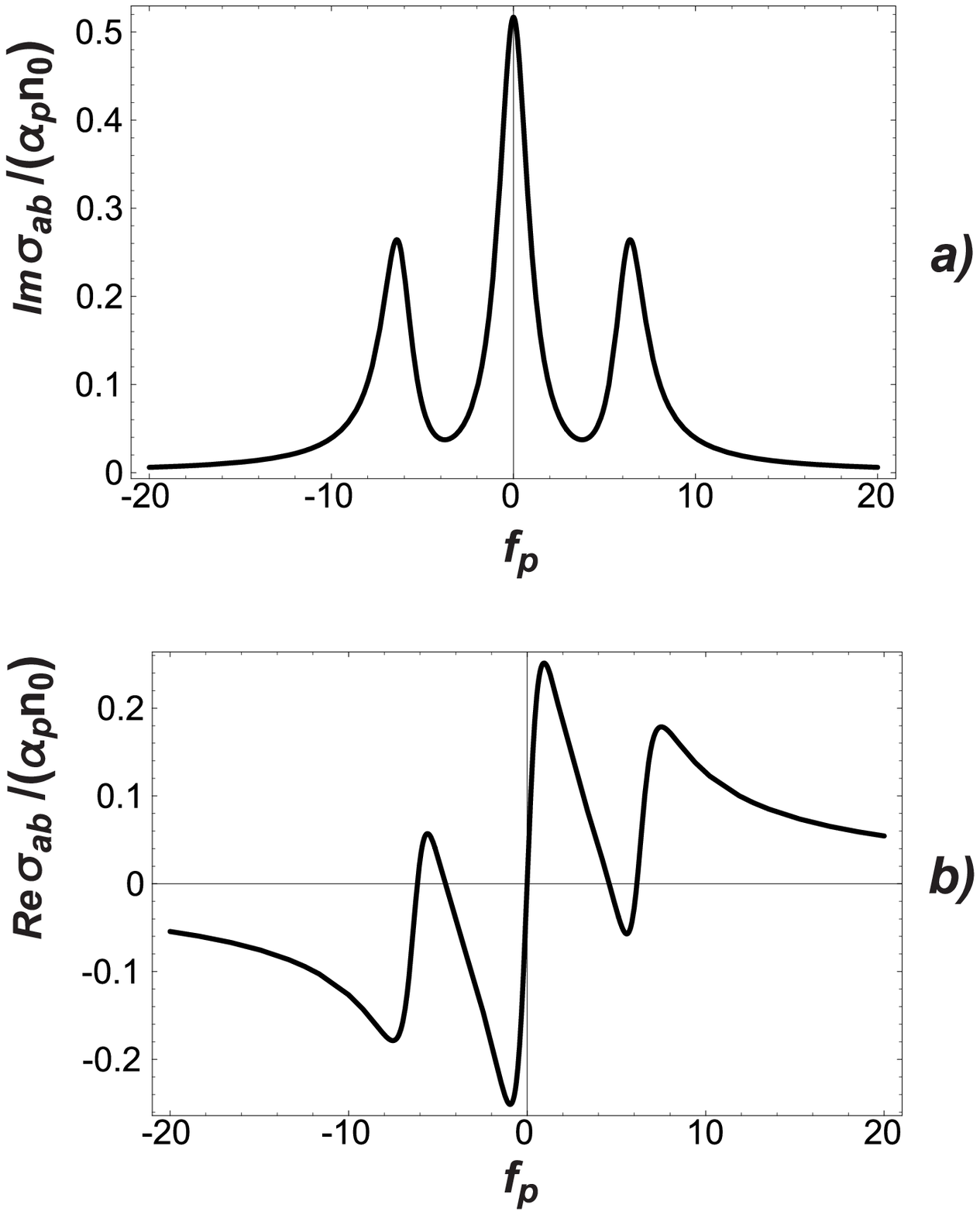}
 \caption{Frequency dependence of the imaginary $a)$ and real $b)$ parts of $\sigma_{ab}
 $ with $h=0$ and $|\alpha|^2=10$ for the immovable four-level atoms
 }
 \label{fig3}
 \end{figure}

 \begin{figure}[t]
  \includegraphics[width=120mm]{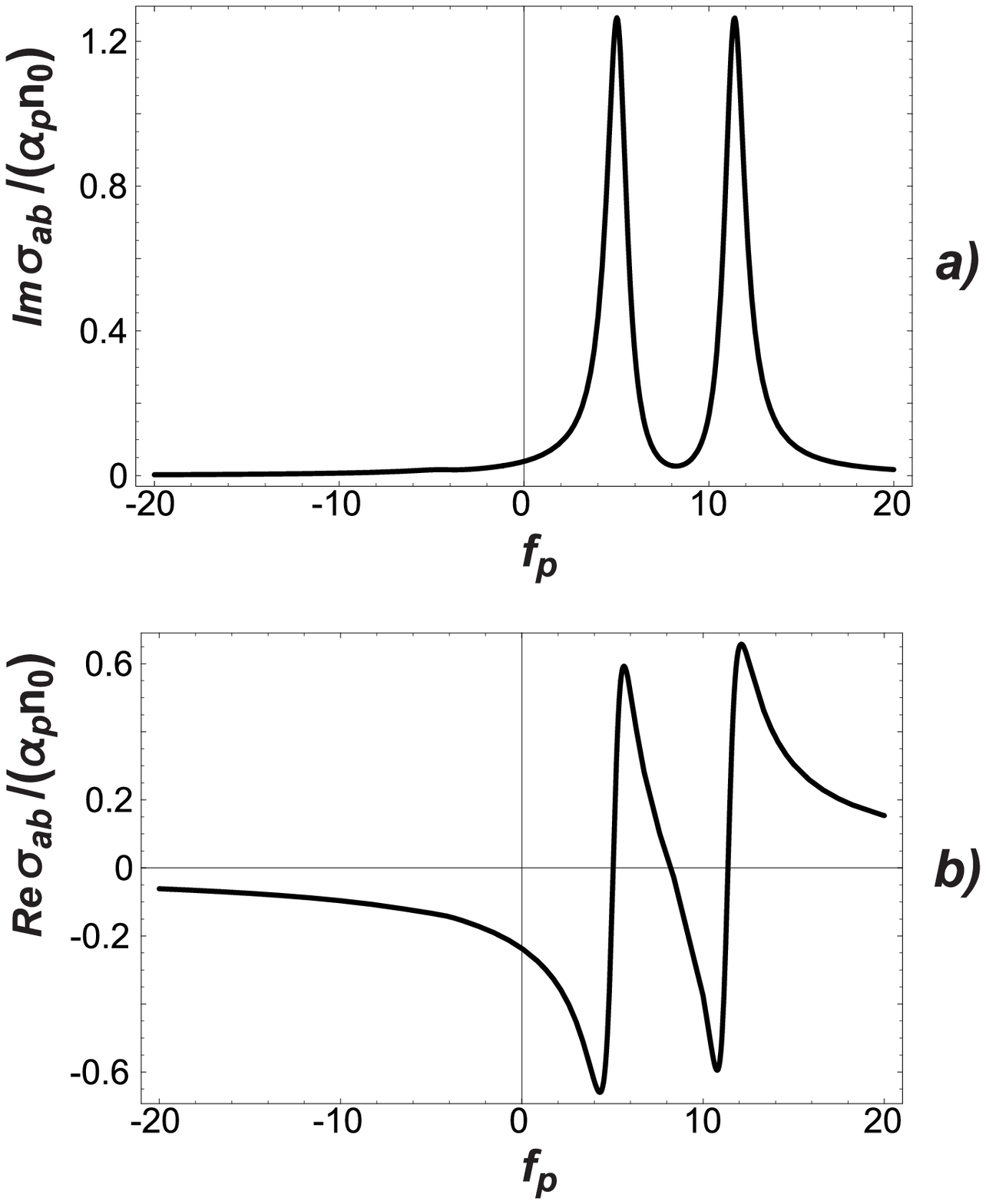}
 \caption{Frequency dependence of the imaginary $a)$ and real $b)$ parts of $\sigma_{ab}
 $ with $h=7$ and $|\alpha|^2=10$ for the immovable four-level atoms}
 \label{fig4}
 \end{figure}

 \begin{figure}[t]
 \includegraphics[width=120mm]{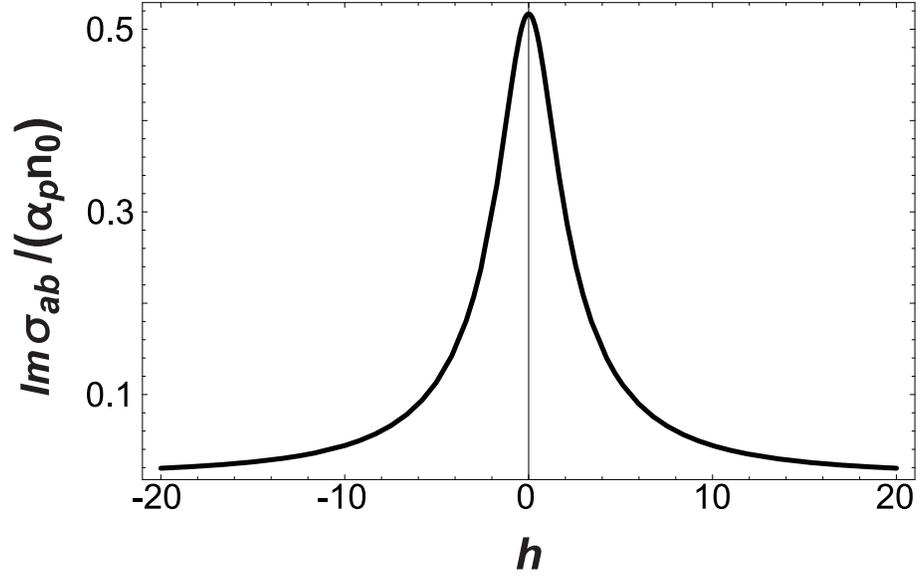}
 \caption{Dependence of the absorption of the immovable atoms on the magnetic field $h$
  in the EIT-EIA area $f_p=h$}
 \label{fig5}
 \end{figure}

\begin{figure}[t]
 \includegraphics[width=120mm]{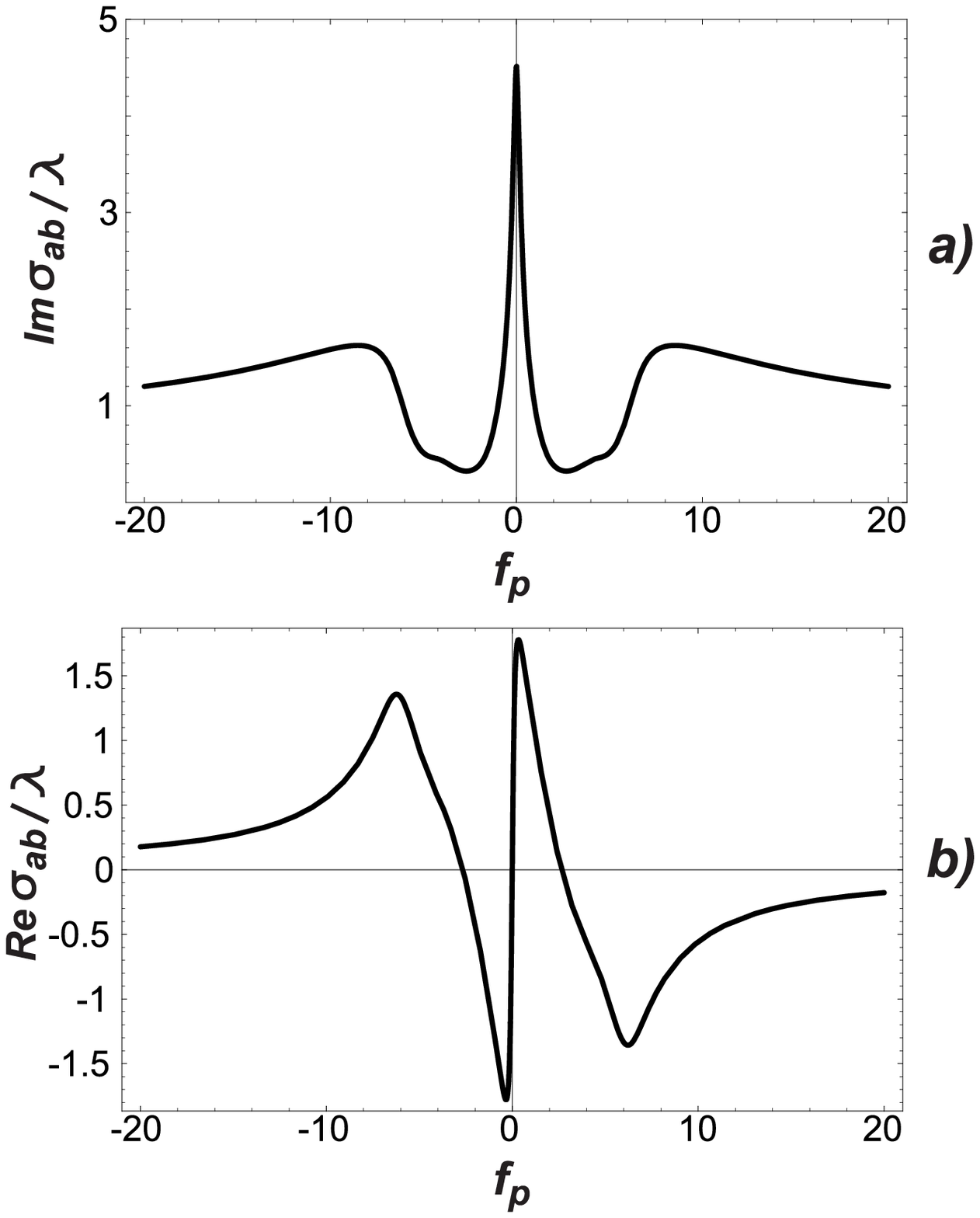}
 \caption{Frequency dependence of the imaginary $a)$ and real $b)$ parts of $\sigma_{ab}
 $ with $h=0$ and $|\alpha|^2=10,\;\varepsilon_0=1/10$ for the thermal four-level atoms, $\lambda=\alpha_pn_0\sqrt\pi/x_0$}
 \label{fig6}
 \end{figure}

\begin{figure}[t]
 \includegraphics[width=120mm]{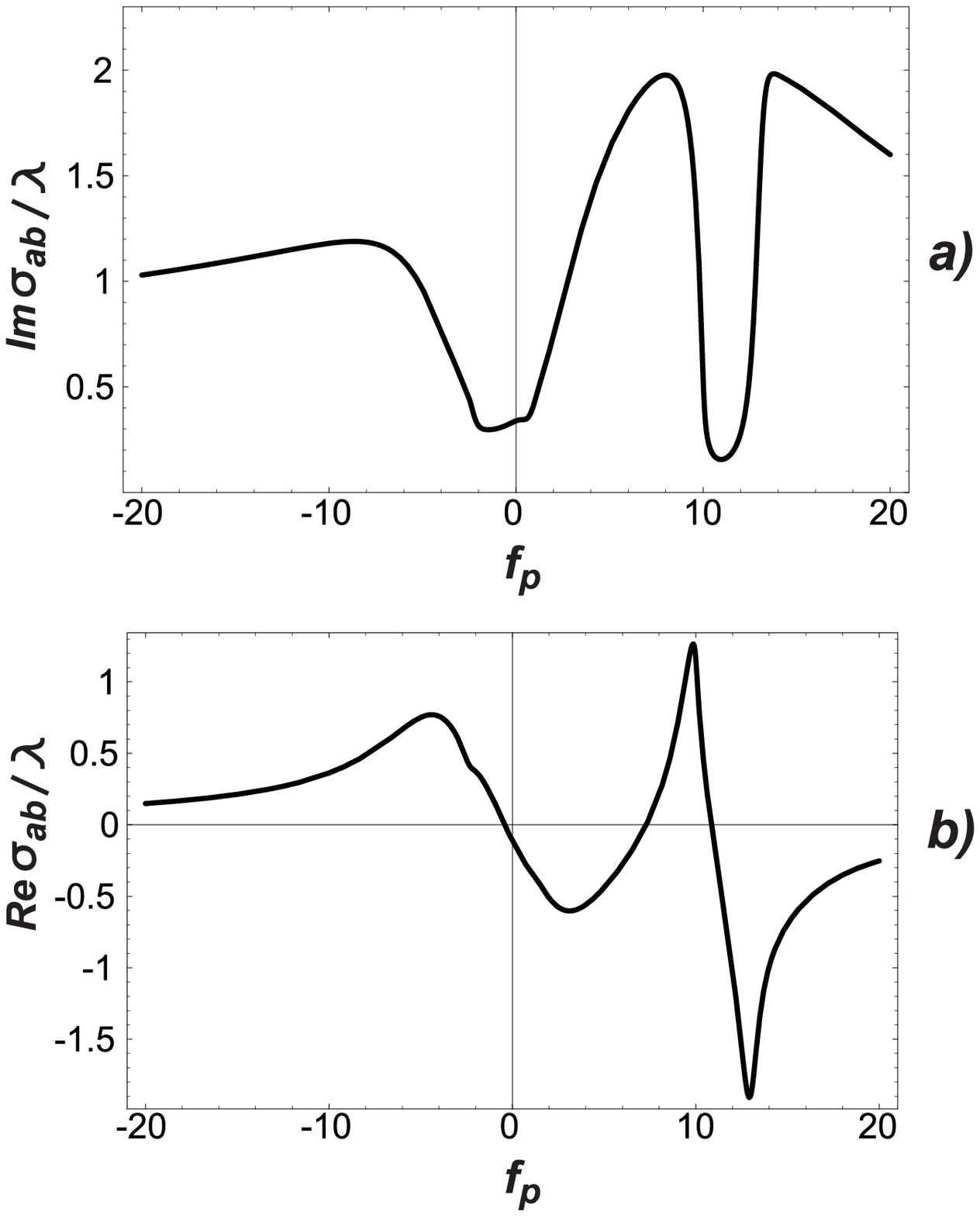}
 \caption{Frequency dependence of the imaginary $a)$ and real $b)$ parts of $\sigma_{ab}
 $ with $h=10$ and $|\alpha|^2=10,\;\varepsilon_0=1/10$ for the thermal four-level atoms}
 \label{fig7}
 \end{figure}

\begin{figure}[t]
 \includegraphics[width=120mm]{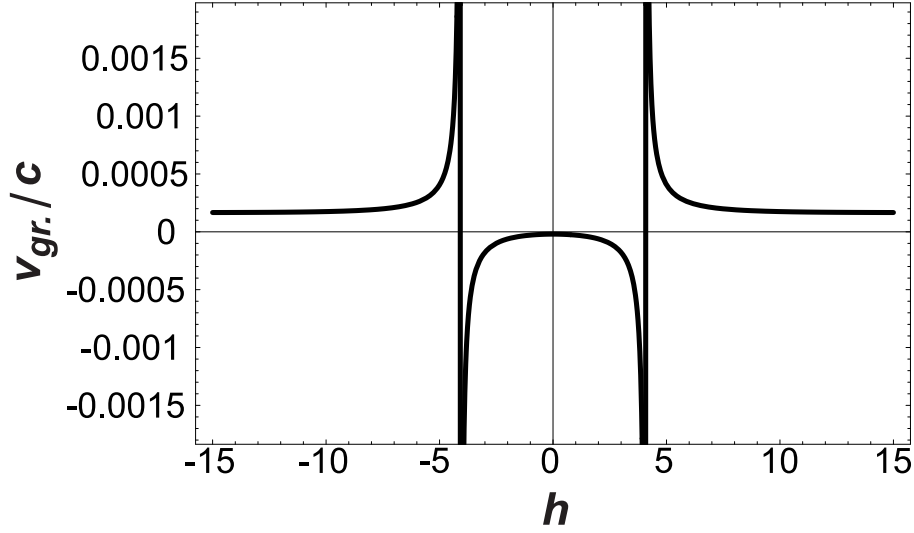}
 \caption{Dependence of the group velocity on the power
 of the externally applied magnetic field in the EIT/EIA area ($f_p=h$) with
 $|\alpha|^2=10,\;\varepsilon_0=1/10$ }
 \label{fig8}
 \end{figure}

\begin{figure}[t]
 \includegraphics[width=120mm]{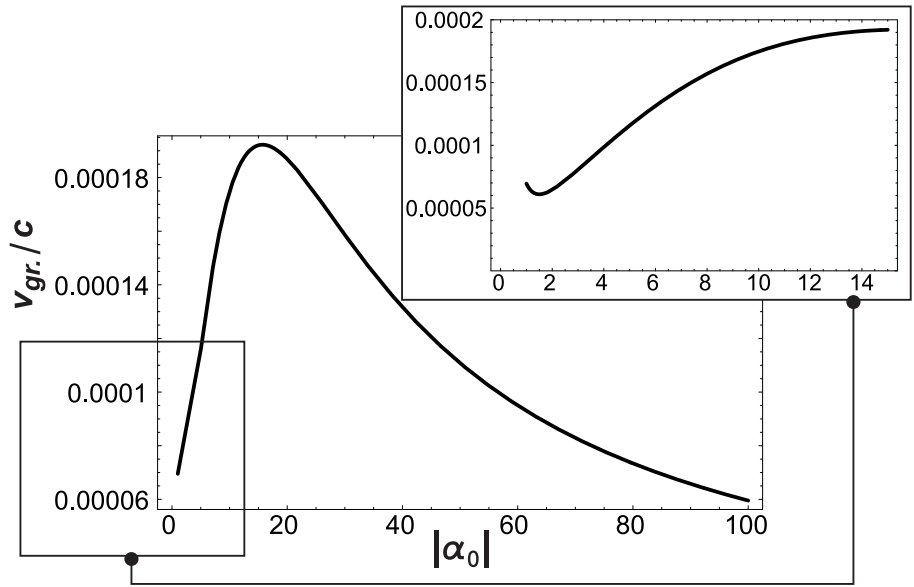}
 \caption{Dependence of the group velocity on the power
 of the drive field in the EIT area ($f_p=h=10$) with
 $\varepsilon_0=1/10$}
 \label{fig9}
 \end{figure}

\end{document}